\documentclass{emulateapj}
\usepackage{./apjfonts}

\usepackage{epsfig}

\def\P3hat{{\mathaccent 94 P}_3}

\def\nancay{Nan\c{c}ay}

\def\qo{{\it q_{\circ}}}
\def\qi{{\it q_{i}}}

\def\viss{{V_{\rm iss}}}
\def\vlos{{V_{\perp}}}
\def\tdiff{{T_{\rm diff}}}
\def\tausc{{\tau_{\rm sc}}}
\def\tref{{T_{\rm ref}}}
\def\dmt{{DM(t)}}

\shorttitle{Timing PSR B1937+21 -- Interstellar Plasma Weather}

\shortauthors{R. Ramachandran et al.}

\begin{document}

\title{Interstellar Plasma Weather Effects in Long-term
Multi-frequency \\ 
Timing of Pulsar B1937+21}

\author{R. Ramachandran, P. Demorest, D. C. Backer}
\affil{Department of Astronomy and Radio Astronomy Laboratory, 
    University of California, Berkeley, CA 94720-3411, USA; \\ e-mail: 
       ramach, demorest, dbacker@astro.berkeley.edu}
\author{I. Cognard} 
\affil{Laboratoire de Physique et Chimie
       de l'Environnement, CNRS, 3A avenue de la Recherche
       Scientifique, F-45071 Orleans, France}
\author{A. Lommen}
\affil{Department of Physics and Astronomy, Franklin and Marshall College, 
       P.O.Box 3003, Lancaster, PA 17604, USA}

\begin{abstract}
We report here on variable propagation effects in over twenty years of
multi-frequency timing analysis of pulsar PSR B1937+21 that determine
small-scale properties of the intervening plasma as it drifts through
the sight line. The phase structure function derived from the
dispersion measure variations is in remarkable agreement with that
expected from the Kolmogorov spectrum, with a power law index of
$3.66\pm 0.04$, valid over an inferred scale range of 0.2---50 A.U.
The observed flux variation time scale and the modulation index, along
with their frequency dependence, are discrepant with the values
expected from a Kolmogorov spectrum with infinitismally small inner
scale cutoff, suggesting a caustic-dominated regime of interstellar
optics. This implies an inner scale cutoff to the spectrum of $\sim
1.3\times 10^9$ meters. Our timing solutions indicate a transverse
velocity of 9 km sec$^{-1}$ with respect to the solar system
barycenter, and 80 km sec$^{-1}$ with respect to the pulsar's LSR. We
interpret the frequency dependent variations of DM as a result of the
apparent angular broadening of the source, which is a sensitive
function of frequency ($\propto\nu^{-2.2}$). The error introduced by
this in timing this pulsar is $\sim$2.2 $\mu$s at 1 GHz. The timing
error introduced by ``image wandering'' from the slow, nominally
refractive scintillation effects is about 125 nanosec at 1 GHz. The
error accumulated due to positional error (due to image wandering) in
solar system barycentric corrections is about 85 nanosec at 1 GHz.
\end{abstract}

\keywords{ISM: general --- pulsars: general --- radio continuum:
general --- scattering --- turbulence}

\section{Introduction}
The dispersion measure (DM) of a pulsar probes
the column density of free electrons along the line of sight (LOS). 
Observed DM variations over time scales of several weeks to years
sample structures in the electron plasma over length scales
of $10^{10}$ m to $10^{12}$ m. Diffraction of pulsar signals is the
result of scattering by
structures on scales below the Fresnel radius, $10^{8}$ m or so.
The DM as well as the scattering measure (SM) variability along the LOS
to the Crab pulsar was first reported by Rankin \& Isaacman (1977),
who reported that the DM variability poorly correlated with the SM
variability. Helfand et al. (1980) inferred an upper limit for DM
variations of a few parts in a thousand for several pulsars.  In an
earlier study of PSR B1937+21 Cordes et al. (1990) measured a DM
change of $\Delta DM\sim 0.003$ pc cm$^{-3}$ over a period of a
thousand days. The work of Phillips \& Wolszczan (1991) presented the
variations of DM observed along the LOS to a few pulsars. They
connected these variations to those on diffractive scales, 
and derived an electron density fluctuation
spectrum slope of $3.85\pm0.04$ over a scale range of $10^7-10^{13}$
meters. Backer et al. (1993) report on further DM variability and
show that the amplitude of the variations known at that time are
consistent with a scaling by the square root of DM.  Another important
investigation by Kaspi et al. (1994) studied DM variations of the
millisecond pulsars PSR B1937+21 and B1855+09 over a time interval of
calendar years 1984 to 1993. In addition to establishing a secular
variation in DM over this time interval, they also show that the
underlying density power spectrum has an index of 3.874$\pm$0.011,
which is close to what we would expect if the density fluctuations are
described by Kolmogorov spectrum. An anomalous dispersion event
towards the Crab pulsar was reported by Backer et al. (2000), where
they report a DM ``jump'' as large as 0.1 pc cm$^{-3}$.

\begin{figure}
\begin{center}
\epsfig{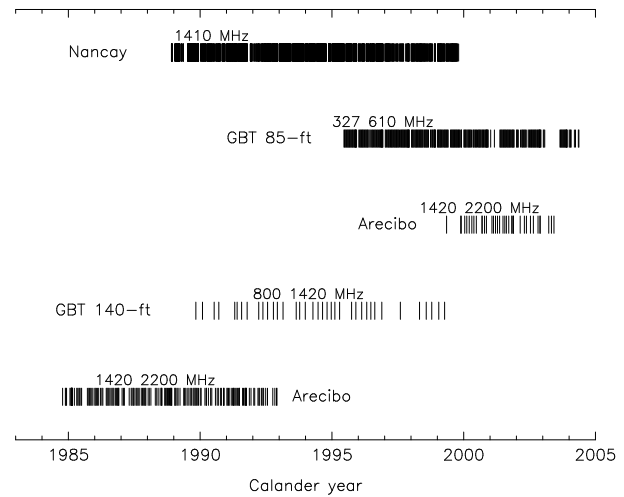}
\caption[]{Summary of our data sample. See text for details.}
\label{fig:obs_summary}
\end{center}
\end{figure}

In this work, we present results of various
long term monitoring programs on PSR
B1937+21. Our data, which includes that of Kaspi, Taylor \& Ryba
(1994), spans calendar years from 1983 to 2004. These
data sets have been taken with five different telescopes, the 
NRAO\footnote{The National Radio Astronomy Observatory (NRAO) 
is owned and operated by Associated Universities, Inc under 
contract with the US National Science Foundation.} Green
Bank 42-m (140-ft) and 26-m (85-ft) telescopes, the 
NAIC\footnote{The National Astronomy and Ionosphere Center is operated
by Cornell University under contract with the US National Science 
Foundation.} Arecibo telescope and the \nancay ~telescope, at 
frequency bands of 327, 610, 800, 1400 and 2200 MHz. 
After giving the details of our observations in \S\ref{sec-obsvn}, 
we describe our analysis methods in \S\ref{sec-anal}. This is 
followed in \S\ref{sec-sightline} by a discussion on the distribution of 
scattering material along the LOS. As we describe, the knowledge 
of temporal and angular broadening of the source, proper motion, and 
scintillation based velocity estimates enables us to at least qualitatively
study the distribution of scattering matter as well as properties
of its wave-number spectrum.

We have measured some of the basic refractive scintillation parameters from 
our observations, and these are discussed in \S\ref{sec-diffref}. 
The frequency dependence of the refractive scintillation 
time scale and the modulation index indicate a caustic-dominated
regime that results from a large inner scale in the spectrum.

We have detected DM variations as a function of time and frequency. 
We determine the phase structure function of the medium with the 
knowledge of the time dependent DM variations, which is consistent 
with a Kolmogorov distribution of density 
fluctuations between scale sizes of about 1 to 100 A.U. These are 
summarized in \S\ref{sec-dmvar} and \S\ref{sec-dmfreq}.

\begin{table}
\begin{center}
\caption[]{Template fit parameters at various frequencies.}
\begin{tabular}{lcccccccc} \hline
{\bf $\nu$ (MHz)} & backend & ~~$w_1$~~ & ~~$l_2$~~ 
& ~~$w_2$~~ & ~~$h_2$~~ & ~~$l_3$~~ & ~~$w_3$~~ & ~~$h_3$~~ \\ \hline\hline
327	 & GBPP	& 8.7 & 0	& 0 & 0 & 186.9  & 12.0 & 0.51 \\
430	 & ABPP	& 10.3	& 7.0	& 2.5	& 0.19	& 186.8	& 12.4	& 0.53 \\
610	 & GBPP	& 9.4	& 7.1	& 2.9	& 0.34	& 187.3	& 10.8	& 0.60 \\
863	 & EBPP	& 8.9	& 7.7	& 5.2	& 0.38	& 187.7	& 10.9	& 0.55 \\
1000 & GBPP & 9.2   & 8.6   & 3.9   & 0.56  & 187.9 & 11.3  & 0.54 \\
1419 & ABPP	& 8.5	& 8.5	& 3.7	& 0.37	& 187.5	& 10.1	& 0.45 \\
1689 & EBPP	& 9.1	& 9.2	& 3.9	& 0.37	& 187.4	& 10.5	& 0.40 \\
2200 & GBPP	& 9.4	& 9.8	& 3.2	& 0.29	& 187.6	& 10.4	& 0.36 \\
2379 & ABPP	& 10.0	& 9.8	& 3.0	& 0.29	& 187.1	& 10.8	& 0.33 \\
\hline
\end{tabular}
\label{tab:fitpars}
\end{center}
\end{table}

PSR B1937+21 is known for its short term timing stability. However, 
the achievable long term timing accuracy is suspected to be 
seriously limited by the interstellar scattering properties. With
our sensitive measurements, we are in a position to quantify these 
errors. In \S\ref{sec-timingerror}, we describe in detail various 
sources of these errors and quantify them.

\section{Observations }
\label{sec-obsvn}
We have used five different primary data sets for this analysis. 
The first set is the 1984--1992 Arecibo pulse timing and
dispersion measurements obtained by Kaspi et al. (1994; hereafter KTR94). 
Their observations were performed with their Mark~II backend (Rawley
1986; Rawley, Taylor \& Davis 1988) and later their Mark~III  backend 
(Stinebring et al. 1992) at two different radio frequency
bands, 1420 MHz and 2200 MHz. Their analysis methods are described in
KTR94.

The second data set is from 800-MHz and 1400-MHz observations 
at the NRAO 140-ft telescope in Green Bank, WV. The Spectral
Processor backend, a hardware FFT device, was used. Details about
the observations and analysis are contained in an earlier report
on dispersion measure variability (Backer et al. 1993).

The third data set consists of observations at 327 MHz and 610 MHz
using the 26m (85-ft) pulsar monitoring telescope at NRAO's Green
Bank, WV site.  Room temperature (uncooled) receivers at the two bands
are mounted off-axis. At 327 MHz the total bandwidth used was 5.5 MHz,
and 16 MHz was used at 610 MHz. The two orthogonally polarized signals
were split into 32 frequency channels in a hybrid analog/digital
filter bank in the GBPP (Green Bank--Berkeley Pulsar Processor).
Dispersion effects were removed in the GBPP in real-time with a
coherent (voltage) deconvolution algorithm.  At the end of the
real-time processing folded pulse profiles were recorded for each
frequency channel and polarization.  Further details of the backend
and analysis can be found in Backer, Wong \& Valanju (2000).  PSR
B1937+21 was observed for about two hours per day starting in
mid-1995.

The fourth data set comes from a bi-monthly precision timing program
that includes B1937+21 at the Arecibo Observatory which we started in
1999 after the telescope upgrade. Signals at 1420 MHz and 2200 MHz
were recorded using the ABPP backend (Arecibo--Berkeley Pulsar
Processor), which is identical to the GBPP.  Our typical observing
sessions at 1420 MHz and 2200 MHz had bandwidths of 45 MHz and 56 MHz,
respectively, and integration times of approximately 10 minutes per
session.

\begin{table}
\tt
\begin{center}
\caption[]{Parameters of PSR B1937+21.}
\begin{tabular}{ll} \hline
{\bf Parameter} & {\bf value} \\ \hline\hline
PSR  & 1937+21 \\
RAJ (hh:mm:ss) & 19:39:38.561 (1) \\
DECJ (dd:mm:ss) & 21:34:59.136 (6) \\
PMRA (mas yr$^{-1}$) & 0.04 (20) \\
PMDEC (mas yr$^{-1}$) & -0.45 (6) \\
$f$ (Hz) & 641.9282626021 (1) \\
$\dot{f}_{-15}$ (Hz s$^{-1}$) & -43.3170 (6) \\
$\ddot{f}_{-26}$ (Hz s$^{-2}$) & 1.5 (3) \\
PEPOCH (MJD) & 47500.000000 \\
START        & 45985.943 \\
FINISH       & 52795.286 \\
EPHEM  & DE405 \\
CLK    & UTC (NIST) \\ \hline
\end{tabular}
\label{tab:1937par}
\end{center}
\end{table}

The fifth data set is from a pulsar timing program that has been
ongoing since 1989 October with the large decimetric radio telescope
located at \nancay, France.  The Nan\c cay telescope has a surface
area of 7000 m$^2$, which provides a telescope gain of 1.6 K
Jy$^{-1}$. Observations are performed with dual-linear feeds at
frequencies 1280, 1680 and 1700 MHz. Then the signal is dedispersed by
using a swept frequency oscillator (at 80 MHz) in the receiver IF
chain. The pulse spectra are produced by a digital autocorrelator
with a frequency resolution of 6.25 kHz. Cognard et al. (1995)
describe in detail the backend setup and the analysis procedure.

A small amount of additional data from the Effelsberg telescope was
used in our profile analysis. At Effelsberg the EBPP backend, a copy
of the GBPP/ABPP, was used.

\section{Basic analysis }
\label{sec-anal}

We first present several results from the analysis of these data
sets: a description of the frequency-dependent profile template used
for timing; spin and astrometric timing parameters from high frequency
data; pulse broadening, flux densities and dispersion measure
as functions of time. In  \S\ref{sec-sightline} we proceed to interpret
these results and return to finer details regarding dispersion measure
variations in \S\ref{sec-dmvar}.

Our basic data set consists of average pulse profiles obtained
approximately every 5 minutes in each of the radio frequency bands --
327, 610, 800, 1420 and 2200 MHz. Figure \ref{fig:obs_summary} provides
a graphical summary of observation epochs vs date.
For data sets corresponding to all
frequencies except 327 MHz, {\it Times of Arrival} (TOAs) were
computed by cross correlating these average profiles with a template
profile. The template profile at a given frequency was made by using
multiple Gaussian fits to very high signal to noise ratio average profiles
at that frequency; the interactive program {\it bfit}, which is 
based on M. Kramer's original program {\it fit} was used.
These fit parameters are listed in Table 1.
Col.~1 in the Table gives the radio frequency
and the backend name is in col.~2. Col.~3 gives the width
of component 1 ($w_1$; its location is taken to be 0 degrees and its
amplitude is set to 1.0); cols.~4-6 and cols.~7-9 give the location ($l$),
width ($w$) and amplitude ($h$) values for components 2 and 3,
respectively. The location and width are given in units of
longitudinal degrees, where 360$^{\circ}$ indicates one full rotation
cycle. The results of this analysis can be compared with that of
Foster et al. (1991) which are given on the line at 1000 MHz
\footnote{The widths $w_1$ and $w_3$ are inverted in Table 4 of Foster
et al.}. There is reasonable agreement for all values except $h_2$
which must have been erroneously entered in Table 4 of Foster et al.
In our analysis templates corresponding to arbitrary frequencies are 
produced by spline-interpolation of the component parameters.

We used the Arecibo (1420 MHz and 2200 MHz) TOAs, and the GBT 140-ft
(800 MHz and 1420 MHz) TOAs to fit for pulsar spin (rotation frequency
($f$), first time derivative ($\dot{f}$), and second time derivative
($\ddot{f}$)) and astrometric (position ({\tt RAJ}, {\tt DECJ}),
proper motion ({\tt PMRA}, $\mu_{\alpha}$, along right acension, and
{\tt PMDEC}, $\mu_{\delta}$, along declination) parameters. All TOAs
were referred to the UTC time scale kept by the National Institute of
Standards and Technology (NIST) via GPS satellite comparison. We
removed the effects of variable dispersion from this fitting procedure
with weekly estimation of DMs and subsequent extrapolation of the dual
frequency data to infinite frequency prior to parameter
estimation. The nature of achromatic timing noise makes it
particularly difficult to determine a precise timing model. As one
adds additional higher derivatives of rotation frequency (e.g., a
third derivative), the best fit parameters change by amounts much
larger than the nominal errors reported by the package that we used,
TEMPO. The results are listed in Table 2. The errors
presented in the table incorporates the range of variation of each
parameter, as additional derivative terms are included. In comparison
to Kaspi, Taylor \& Ryba (1994), the derived proper motion values are
marginally different. We attribute this difference to the variable
influence of timing noise. An important point that needs to be
stressed here is that there is no reason for us to assume that the
higher derivative terms of rotation period (e.g., $\ddot{f}$ or
higher) has anything to do with the radiative braking index. They are
most likely dominated by some intrinsic instabilities of the star
itself, or some other perturbation on the star.

\begin{figure}
\begin{center}
\epsfig{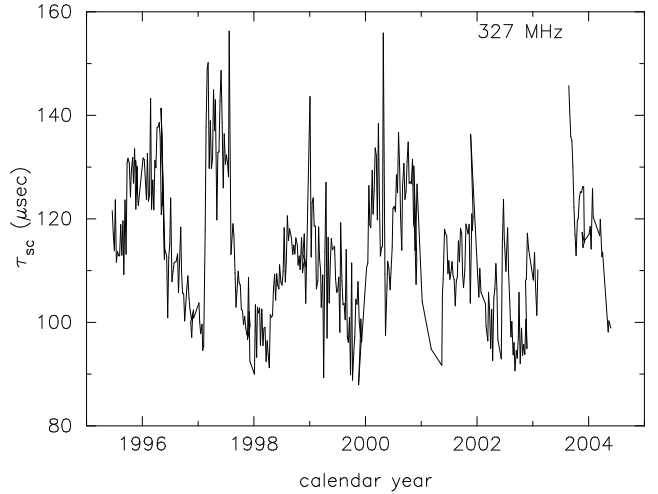}
\caption[]{Measured temporal pulse broadening timescale ($\tausc$) as
a function of time at 327 MHz.}
\label{fig:tausc}
\end{center}
\end{figure}

Extension of dispersion measurement to 327 MHz requires removal of the
time-variable broadening of the intrinsic pulse profile owing to
multipath propagation in the interstellar medium. We deconvolved the
effect of interstellar scattering following precepts first introduced
by Rankin et al. (1970). We assume that the interstellar temporal
broadening is quantified in terms of convolution of a Gaussian
function and a truncated exponential function. If there is only one
scattering screen along the LOS, the assumption of a truncated
exponential function will suffice to represent the scatter
broadening. However, since the scattering may arise from material
distributed all along the LOS, a more realistic representation is
approximated by a truncated exponential function ``smoothed''
(convolved) with a Gaussian function. The intrinsic pulse profile was
estimated by extrapolation of parameters from the higher frequency
profiles.  In the deconvolution procedure, we minimized the normalized
$\chi^2$ value by varying the width of the Gaussian $w_g$ and the
decay time scale of the truncated exponential $\tau_e$, while keeping
the intrinsic pulse profile fixed. The pulse scatter broadening is
quantified as $\tausc\; = \; (w_g^2 + \tau_e^2)^{1/2}$. We repeated this 
for average profiles obtained at every epoch to obtain the $\tausc$ 
measurement. In our
fits, the average value of $w_g$ came to about 74 $\mu$sec,
whereas the corresponding value for $\tau_e$ was about 85 $\mu$sec.
The measurement of $\tausc$ versus time at 327~MHz is plotted in the
Figure \ref{fig:tausc}. This quantity has a mean value of 120~$\mu$s,
an RMS variation of 20~$\mu$s, and a fluctuation timescale of
$\sim60$~days. We explain these variations as the result of refractive
modulation of this inherently diffractive parameter in discussion
below.  The estimated RMS variation at the next higher frequency in
our data set, 610 MHz, is $\sim$2.5 $\mu$s, using a frequency
dependence of $\tausc\propto\nu^{-4.4}$.  This is too small to allow
fitting at this frequency band.

\begin{figure*}
\begin{center}
\epsfig{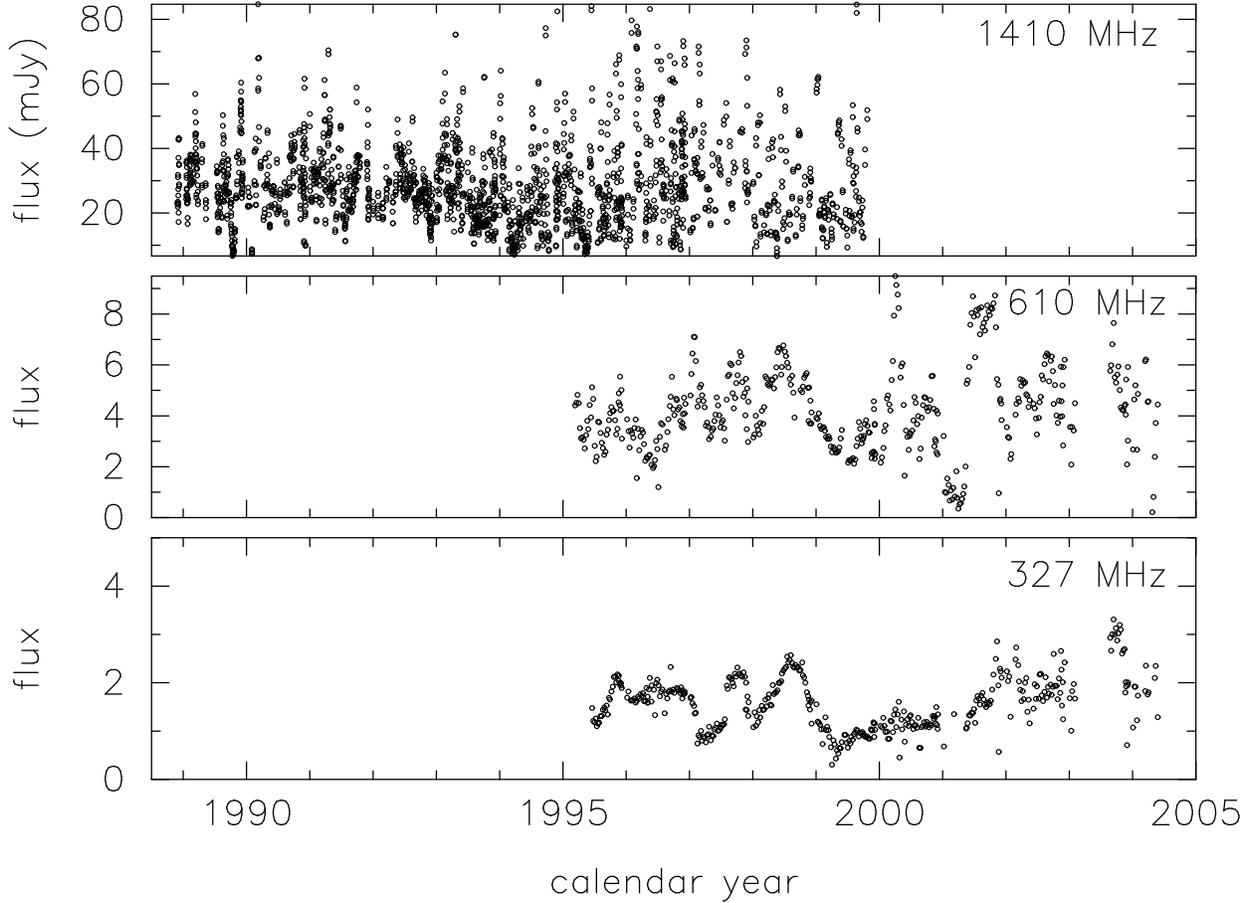}
\caption[]{Measured flux density as a function of time. The top panel
corresponds to the radio frequency of 1410 MHz with the data obtained
from \nancay ~telescope, the middle and the bottom panel to 610 and
327 MHz with the data obtained from the Green Bank 85-ft telescope.}
\label{fig:fluxplot}
\end{center}
\end{figure*}

In the strong scintillation regime, time dependent variations in the
observed flux occur in two distinct regimes --- {\it diffractive}
and {\it refractive}. The diffractive effects are dominated by
structures smaller than the Fresnel scale, and appear on short time
scales and over narrow bandwidths. In our observations diffractive
modulations are strongly suppressed. On the other hand refractive
effects occur on days time scales and are correlated over wide
bandwidths. We have analyzed our best data sets -- the densely sampled
data at 327 MHz and 610 MHz from Green Bank and at 1410 MHz from Nan\c
cay -- for flux density variations as a function of time. The data
are presented in Figure \ref{fig:fluxplot}.

In analyzing the low frequency flux data from Green Bank, we have not
adopted a rigorous flux calibration procedure.  While there is a pulsed
calibration noise source installed in this system, equipment changes and
the nature of the automated observing have led to large gaps in the
calibration record.  Rather than dealing with a mix of calibrated and
uncalibrated data, or lose a large fraction of the data, we decided not
to apply any calibration.  Instead, we normalize our data by assuming
the system temperature is constant.  In order to see what effect this
has on our results, we did two tests.

First, we analyzed observations of PSR~B1641$-$45, taken with the same
system, over a similar time range.  This pulsar is known to have a very
long refractive timescale, $\tref >1800$~days (Kaspi \& Stinebring,
1992), so it can be used as a flux calibrator.  In our data, we
find it to have a modulation index, $m=0.10$.  This immediately puts a
upper limit of 10\% on any systematic gain and/or system
temperature variations.  Since modulation adds in
quadrature, and we observe modulation indeces of $m\sim0.4$ for
PSR~B1937+21, gain fluctuations represent at most a small fraction
of the observed modulation.

We also considered the possibility that gain variations could influence
our measurement of $\tref $.  This might happen if they occur with a
characteristic timescale longer than 1~day.  In order to test this, we
analyzed observations of the Crab pulsar, PSR~B0531+21, again taken with
the same system over the same time range.  The refractive parameters of
this pulsar were studied in detail by Rickett \& Lyne (1990).  It makes
a good comparison since it has modulation index of $m=0.4$ at 610~MHz,
very similar to PSR~B1937+21.  Applying the structure function analysis
(see \S\ref{sec-diffref}) to this data gives $\tref =11$~days at
610~MHz, and $\tref =63$~days at 327~MHz, consistent with the previously
published results and a scaling law of $\tref \propto\nu^{-2.2}$.

The procedure that we have adopted to calibrate our data set from
\nancay~ telescope is described in detail in Cognard et al. (1995).

\section{Distribution of scattering material along the line of sight}
\label{sec-sightline}

Several authors have shown how the scattering parameters of a pulsar
can be used to assess the distribution of scattering material along
the LOS (Gwinn et al. 1993; Deshpande \& Ramachandran 1998; Cordes \&
Rickett 1998).  This results from the varied dependences of the
scattering parameters on the fractional distance of scattering
material along the LOS. PSR B1937+21 is viewed through the local
spiral arm as well as the Sagittarius arm which are both potential
sites of strong scattering.  The parameters employed in this analysis
are: the temporal pulse broadening by scattering ($\tausc$; or its
conjugate parameter $\Delta\nu$, the diffractive scintillation
bandwidth), the diffractive scintillation time scale ($\tdiff$), the
angular broadening from scattering ($\theta_H$), the proper motion of
the pulsar ($\mu_{\alpha}$, $\mu_{\delta}$), and a distance estimate
of the pulsar ($D$).

Let us first compare $\theta_H$ and $\tausc$ that are the result of
multiple scattering along the LOS, and express them as
(Blandford \& Narayan 1985)

\begin{eqnarray} 
\tausc &=& \frac{1}{2cD}\int_0^D x(D-x)\;\psi(x)\; {\rm d}x \\
\theta_H^2 &=& \frac{4\ln2}{D^2}\int_0^D x^2\psi(x)\; {\rm d}x.
\label{eq:tautheta}
\end{eqnarray}

In these equations, $x$ is the coordinate along the LOS, with the
pulsar at $x=0$ and the observer at $x=D$.  $\psi(x)$ is the mean
scattering rate. If the scattering material is uniformly distributed
along the LOS, then the relation between the two quantities can be
expressed as $\theta_H^2=16\ln2\left(c\tausc/D\right)$. With the
distance to the pulsar of 3.6~kpc to the pulsar (according to the
distance model of Cordes \& Lazio 2002), and the average pulse
broadening time scale of 120 $\mu$s (from the present work), we obtain
an estimate of the angular broadening, $\theta_{\tau}$, of 12
mas. This is in modest agreement with the measured value of 14.6$\pm
1.8$ mas (Gwinn et al. 1993), given the uncertainty in the distance to the pulsar and the
simple assumption that the scattering material is uniformly
distributed along the LOS.

\begin{figure}
\begin{center}
\epsfig{file=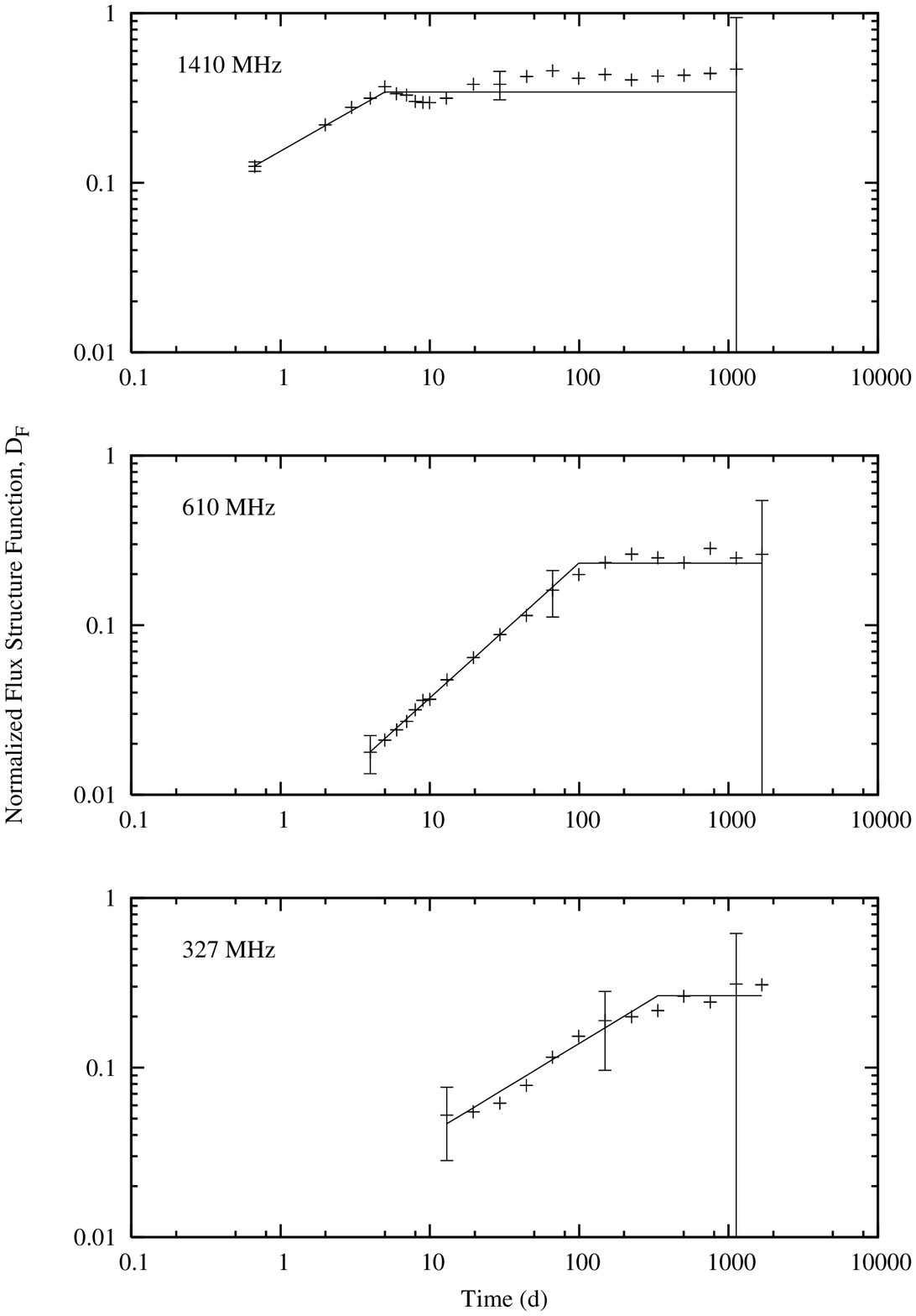,width=9.5cm}
\caption[]{Structure function of normalized flux variations at 1410
MHz ({\it Top}), 610 MHz ({\it middle}), and 327 MHz ({\it
bottom}). The 1410 MHz data was obtained from \nancay telescope. The
saturation value of the structure function at larger lag values
indicates the observed modulation index. Error bars are shown on only
a few points, to preserve clarity. See text for details.}
\label{fig:fluxstruct}
\end{center}
\end{figure}

Next, we formulate two approaches to estimation of the velocity of the
LOS with respect to the scattering medium, and use these approaches to
assess the location and extent of the medium.  The transverse velocity
of the pulsar based on the measured proper motion
(Table 2) an assumed distance of $D=3.6$ kpc (Cordes \&
Lazio 2002) is 9 km s$^{-1}$. This value is the velocity of the pulsar
with respect to the solar system barycenter.  With the assumed ``Flat
Rotation Curve'' linear velocity of the Galaxy of 225 km s$^{-1}$, and
the Sun's peculiar velocity of 15.6 km s$^{-1}$ in the Galactic
coordinate direction of $(l,b) = (48.8^{\circ}, 26.3^{\circ})$ 
(Murray 1986), the
transverse velocity of the pulsar in its LSR ($V_p$) is 80 km
s$^{-1}$.

The {\it scintillation velocity}
($\viss$), which  is an estimate of the velocity of the {\it
diffraction pattern} at the location of the Earth, is 
estimated from the decorrelation bandwidth ($\Delta\nu$) and the
diffractive scintillation time scale ($\tdiff$). 
Gupta et al. (1994) conclude that
\begin{equation}
\viss\;=\;3.85\times 10^4\; \sqrt{\frac{D \; z \; \Delta
\nu}{(1-z)}} \; \frac{1}{\tdiff\;\nu_{\rm GHz}} \;\;\; {\rm km\;s^{-1}}
\end{equation}
where $\nu_{\rm GHz}$ is the observing frequency in GHz, $D$ is in
kpc, $\Delta\nu$ is in MHz, and $\tdiff$ is in seconds.  The variable
$z$ gives the fractional distance to the scattering screen, where
$z=0$ gives the observer's position, and $z=1$ gives the pulsar's
position.  The value of decorrelation bandwidth is computed by the
relation $\Delta \nu \; =\; 1 / (2 \pi \tausc)$. When the effective
scattering screen is midway along the LOS ($z=0.5$), $\viss\;=\;V_p$,
and when the screen is at the location of the pulsar ($z=1.0$), $\viss
= \infty$.
While doing this, an important assumption is that the pulsar proper
motion is dominant over contributions from differential Galactic
motion, solar peculiar velocity, and the Earth's annual orbital
modulation.  In the case of PSR B1937+21, this assumption is not
justified. The effective
scattering screen, which is located somewhere along the LOS, has a
Galactic motion whose component along the LOS direction is different
from that of the pulsar or the Sun. In order to correct for this
effect, let us calculate the LOS velocity across the effective
scattering screen at a fractional distance $z$ from the observer:

\begin{equation}
\vlos\;=\; 3.85\times 10^4\; \frac{\sqrt{D \; z \; (1-z)\; \Delta
\nu}}{\tdiff\;\nu_{\rm GHz}} \;\;\; {\rm km\;s^{-1}}
\label{eq:vlos}
\end{equation}

Then, let us assume that the scattering along the LOS can be
adequately expressed by having a thin screen alone, at a distance of
$D_s=zD$ from the observer. Then, Equation \ref{eq:tautheta} can be
expressed as
\begin{eqnarray}
\tausc &=& \frac{\psi_{\circ}}{2c}\; D\;z \; (1-z) \\
\theta_H^2 &=& 4\;\ln 2\; (1-z)^2 \psi_{\circ}
\end{eqnarray}
Here, $\psi_{\circ}$ gives the mean scattering rate corresponding 
to the effective thin screen. Then, let us express independently the 
transverse velocity of the LOS across the scattering screen as
\begin{eqnarray}
\vec{V}_{\perp}' &=& (1-z)\;\vec{V}_e\; + \; z\vec{V}_p \;-\; \vec{V}_G
(zD\hat{n}) \nonumber\\
&=& \vec{V}_E + zD\vec{\mu} - \vec{V}_G (zD\hat{n}),
\label{eq:vlos2}
\end{eqnarray}
where $V_e$ is Earth's orbital velocity, $V_p$ is the pulsar transverse
velocity in its LSR, and $V_G$ is the transverse velocity contribution from 
the Galactic differential motion to the screen. $V_E$ gives the 
contribution of the Earth's motion on the LOS velocity across the screen.

Equations \ref{eq:vlos} and \ref{eq:vlos2} give two independent
estimates of the line of sight velocity across the effective
scattering screen and therefore allow us to solve for the value of $z$ given
$D$. With $D=3.6$ kpc (Cordes \& Lazio 2002), 
we find $z=0.7$. The LOS velocity is 51 km s$^{-1}$. 
The assumed value of $\tdiff$ is
78 seconds at 327 MHz (scaled from 444 seconds at 1400 MHz of Cordes
et al. 1991), and the value of $\Delta\nu$ is 0.0013 MHz calculated
from $\tausc=120$ $\mu$s.

To summarize, the measured value of $\theta_H=14.6\pm 1.8$ mas and the
estimated value of $\theta_{\tau}$ are consistent with each other,
suggesting a uniformly distributed scattering medium. 
On the other hand, comparison of velocity components, $V_p$,
$\vlos$ and $V_{\rm los}^\prime$ suggest a thin-screen at $z\sim
2/3$. As Deshpande \& Ramachandran (1998) demonstrate, this solution
is equivalent to having a uniformly distributed scattering medium!
Therefore, we conclude that the line of sight to PSR B1937+21 can be
described adequately by a uniformly distributed scattering matter.

\begin{figure}
\begin{center}
\epsfig{file=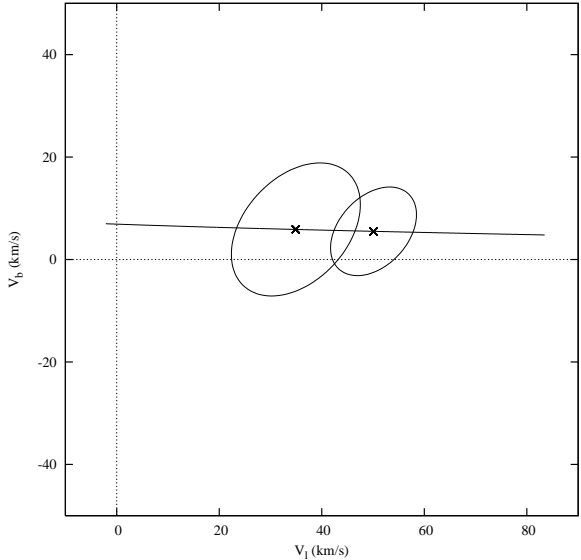,width=8cm}
\caption[]{Estimated line of sight transverse velocity across the
effective scattering screen at a distance of $zD$ from the Sun. The
velocity is resolved into the components along galactic longitude
($V_l$) and latitude ($V_b$). We have assumed a distance of 3.6 kpc to
the pulsar from the Sun. Any point on the line indicates a combination
of $V_l$ and $V_b$ corresponding to a value of $z$, with the left most
end for $z=0$ and the right most end for $z=1$. The line itself does
not include the Earth's orbital velocity contribution. The annual
modulation due to Earth's motion in its orbit is shown as the
ellipses. The two ellipses correspond to the scenarios of $z=0.5$ and
$z=2/3$.}
\label{fig:annualmod}
\end{center}
\end{figure}

The Earth's orbital velocity around the Sun will modulate the observed
scintillation speed, and therefore the diffractive scintillation time
scale, with a one-year time scale. The amplitude of this modulation
will depend on the effective $z$ of the diffracting material, and so
monitoring could provide an estimate of the effective screen location.
If the effective screen is close to the Earth, then the modulation is
strong, and if it is located close to the pulsar, then it is
negligible. Figure \ref{fig:annualmod} demonstrates this effect. The
ordinate and abscissa give the LOS velocity across the effective
scattering screen along the galactic longitude and latitude,
respectively. For an assumed distance of 3.6 kpc, the straight line
shows the expected centroid velocity of $\vlos^\prime$. The left most
end of the line (origin of the plot) corresponds to $z=0$, and the
right most end corresponds to $z=1$. The annual modulation of
$\vlos^\prime$, shown as the two ellipses, correspond to z=0.5 and
$z=2/3$. We have no way of identifying this annual modulation in our
data, as we are insensitive to diffractive effects in our data set.

Another measurement that could help us is the direct measurement of
distance to this object by parallax measurements. Chatterjee et
al. (2005, private communication), from their preliminary Very Long Baseline Array (VLBA)
based parallax measurements, report that the distance to PSR B1937+21
is $2.3^{+0.8}_{-0.5}$ kpc, if they force the proper motion value to
be the same as that of our timing based measurements (Table 2).
In the coming year, accuracy of their measurements
will improve with further sensitive observations.

\section{Refractive scintillation }
\label{sec-diffref}

\subsection{Parameter estimation}
\label{sec-diffrefparm}
We determine refractive scintillation parameters from the data
presented in Figure \ref{fig:fluxplot} following the structure
function approach in previous studies (Stinebring et al 2000; Kaspi \&
Stinebring; Rickett \& Lyne 1990). We define the structure function
$D_F$ for flux time series $F(t)$ as
\begin{equation}
D_F(\delta t) = \frac{\langle [F(t) 
- F(t+\delta t)]^2\rangle}{\langle F(t) \rangle^2},
\label{eq:structflux}
\end{equation}
where $\delta t$ is a time delay. Since our flux measurments occur at 
discrete and unevenly spaced time
interals, we compute the flux difference for all possible lags, then
average results into logarithmically spaced bins.

The flux structure function typically has a form described by Kaspi \&
Stinebring et al. (1992) - a flat, noise dominated section at small
lags, then a power-law increase which finally saturates at a value
$D_s$ at large lags.  In practice, the saturation regime may have
large ripples in it, an effect of the finite length of any data set.
In addition, the measured flux structure function is offset from the
``true'' flux structure function due to a contribution from
uncorrelated measurement errors. At 327~MHz and 610~MHz, we estimate
this noise term from the short-lag (noise regime) values.  At 1410~MHz
(from \nancay), we use the individual flux error bars to get the noise
level.  After subtracting the noise value, we fit the result to a
function of the form
\begin{equation}
D_F(\delta t) = 
    \left\{ \begin{array}{ll} 
        D_s(\delta t/T_s)^\alpha, & [0<\delta t<T_s] \\
        D_s,  & [\delta t > T_s]
    \end{array} \right.
\label{eqn:sffit}
\end{equation}
In this fit, the power law slope $\alpha$, the saturation timescale
$T_s$, and the saturation value $D_s$ are all free parameters. The
flux structure function data and fits are shown in Figure
\ref{fig:fluxstruct}.

As shown in Rickett \& Lyne (1990), the refractive parameters can be
measured from the flux structure function using the following
relationships: The modulation index $m$ is given by $m =
\sqrt{D_s/2}$, and the refractive scintillation timescale $\tref$ is
given by $D_F(\tref) = D_s/2$. All the measured parameters, including
those measured by earlier investigators are summarized in Table 3.


Based on a propagation model through a simple power-law density fluctuation
spectrum, we expect to see refractive variations in the flux measurements 
on a timescale $\tref\sim0.5\theta_H D/\vlos$, where $\vlos$ is the line of
sight velocity across the effective scattering screen.
For the argument sake, if we assume an effective scattering screen at
$z=0.5$, then $\vlos\sim 40$ km sec$^{-1}$.  With $\theta_H = 14.6$
mas, the expected refractive scintillation time scale is $\sim$3 years
at 327 MHz. This is more than an order of magnitude in excess of the
measured value.  Furthermore, if the density fluctuations in the
medium are distributed according to the Kolmogorov power law
distribution, then the expected frequency scaling law is
$\tref\propto\lambda^{2.2}$. Our measured values indicate a
significantly different scaling. Although it is consistent with
$\tref\propto\lambda^{2.2}$ between 610 and 1420 MHz, it is not so
between 327 and 610 MHz, where it is consistent with being directly
proportional to $\lambda$. Our observed modulation index ($m$) values
are also considerably larger than predicted, and show a ``flatter''
wavelength dependence, as listed in Table 3.  We
will address this issue in detail in the following section.

\subsection{Nature of the spectrum -- inner scale cutoff}
\label{sec-cutoff}
The three disagreements with a simple model
summarized in \S\ref{sec-diffrefparm} force us to
explore a few aspects of the electron density power spectrum that may
possibly explain what we observe. The effects of {\it caustics} on the
observed scintillations have been explored by several earlier
investigators, most notably Goodman et al. (1987) and Blandford \&
Narayan (1985). In particular, if the power law scale distribution in
the medium is truncated at an inner scale that is considerably larger
than the diffractive scale, as they show, the observed flux variations
are dominated by caustics. This is of great interest to us, as this
seems to explain all the discrepancies that we note in our observed
refractive parameters. For instance, as Goodman et al. (1987) show
that if the inner scale cutoff is a considerable fraction of the
Fresnel scale, then the observed fluctuation spectrum of flux is
dominated by fluctuation frequencies that are lower than the
diffractive frequencies, but significantly higher than that expected
from refractive scintillation.  This is what we observe.  Moreover, as
they note, the observed wavelength dependence of the refractive time
scale, as well as the modulation index is expected to be ``shallower''
than the expected values of $\lambda^{2.2}$ and $\lambda^{-0.57}$,
respectively.

A ``shallow'' frequency dependence of the modulation index has been
reported by others (Coles et al. 1987; Kaspi \& Stinebring 1992; Gupta
et al. 1993; Stinebring et al. 2000).
While Kaspi \& Stinebring (1992) find that the observed refractive
quantities matched well with the predicted values for five objects,
three other objects, especially PSR B0833--45, has a significantly
shorter measured $\tref$ and greater modulation index than
expected. This is very similar to our situation here with PSR
B1937+21.

Stinebring et al. (2000) concluded that the 21 objects that they
analysed fell into two groups.  The first group followed the frequency
dependence predicted by a Kolmogorov spectrum with the inner cutoff
scale far less than the diffractive scales ({\it Kolmogorov-consistent
group}). The second group, which is the {\it super-Kolmogorov group},
is consistent with a Kolmogorov spectrum with an inner scale cutoff at
$\sim 10^{8}$ meters. The observed modulation indices were
consistently greater than that of the Kolmogorov predictions, as we
have seen in our measurements of PSR B1937+21. This group includes
pulsars like PSRs B0833--45 (Vela), B0531+21 (Crab), B0835--41,
B1911--04 and B1933+16. An important physical property that binds them
all is that, excepting one object, all objects have a strong {\it
thin-screen} scatterer somewhere along the LOS. This is either a
supernova remnant (or a plerion) like in the case of Vela and Crab
pulsars, a HII region as in the case of B1942--03 and B1642--03, or a
Wolf-Reyet star as in the case of B1933+16 (see Prentice \& ter Haar
1969; Smith 1968). Although our measurements show that pulsar PSR
B1937+21 is consistent with the characteristics of the {\it
super-Kolmogorov} group, as we describe in \S\ref{sec-sightline}, we
find no compelling evidence for the presence of any dominant scatterer
somewhere along the LOS.

To summarize, while some investigators have reported agreement of
the measured refractive properties with the theoretical expectations
from a Kolmogorov spectrum with an infinitismally small inner scale,
there are a considerable number of cases where
the observed properties are significantly different from that predicted
by the simple Kolomogorov spectrum.
These other cases can be explained by invoking 
spectrum with a large inner scale cutoff, including the case where
the cutoff approaches the Fresnel radius and leads to
a caustic-dominated regime.
From Gupta et al. (1993) and Stinebring et al. (2000), the modulation
index can be specified as a function of the inner cutoff scale as
\begin{equation}
m\;=\;0.85\; \left( \frac{\Delta\nu}{\nu} \right)^{0.108}\; \left(
\frac{r_i}{10^8{\rm m}} \right)^{0.167} D_{\rm kpc}^{-0.0294}.
\end{equation}
With the known value of $\Delta\nu$ at 327 MHz of 1.33 kHz, the
distance to the pulsar of 3.6 kpc, and the observed modulation index
of 0.39, the inner scale cutoff, $r_i$, comes to $1.3\times 10^9$
meters.

\section{DM variations}
\label{sec-dmvar}

\begin{figure*} 
\begin{center}
\epsfig{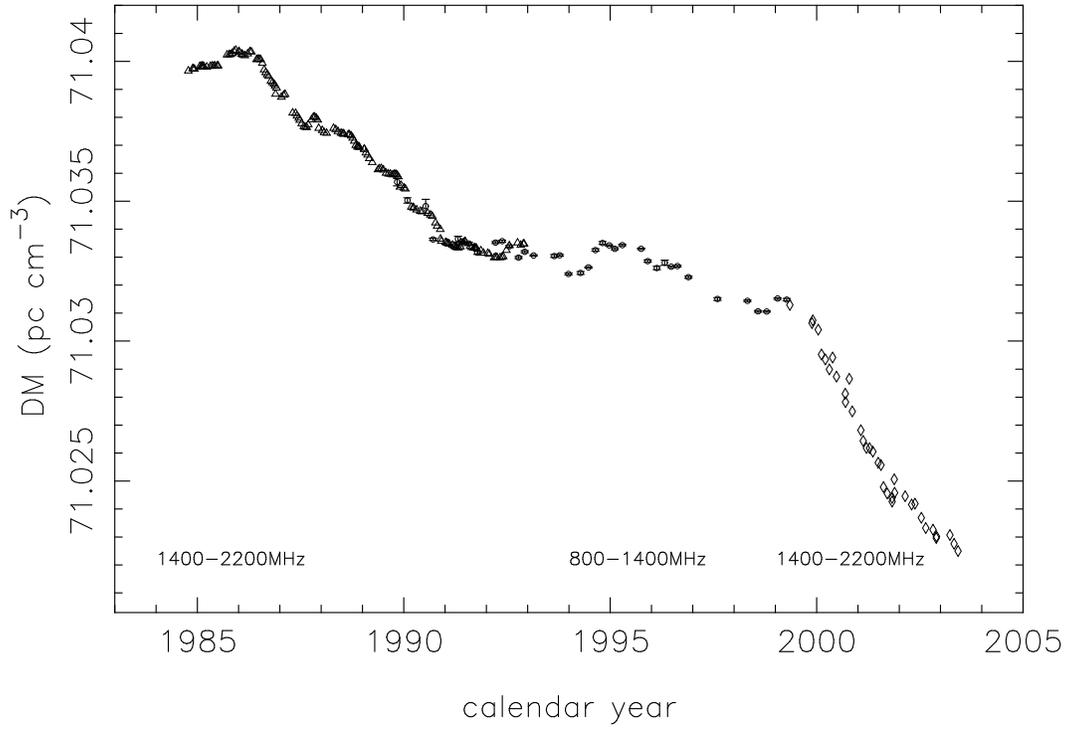}
\caption[]{Dispersion Measure as a function of time. Open triangles give the
measurements of KTR94 at 1400 and 2200 MHz, open circles are from our Green
Bank 140-ft telescope measurements at 800 and 1400 MHz, and the open diamond
symbols indicate our measurements from the Arecibo Observatory, at 1420 and 2200 MHz. All error bars indicate RMS errors.}
\label{fig:dmvstime}
\end{center} 
\end{figure*}

We turn now to the dispersion measure variations presented in Figure
\ref{fig:dmvstime} that sample density variations on transverse scales
much larger than those involved with diffractive and refractive
effects.  The most striking feature in Figure \ref{fig:dmvstime} is
the large secular decline from 71.040 pc cm$^{-3}$ in 1985 to 71.033
pc cm$^{-3}$ in 1991 and then to 71.022 pc cm$^{-3}$ by late
2004. These long-term secular variations are many times greater than
the RMS fluctuations of $\sim 10^{-3}$ pc cm$^{-3}$ on short time
scales.  An important question that arises is whether these variations
are the result of a spectrum of electron-density turbulence, or
whether there might be a contribution from the smooth gradient of a
cloud, or clouds along the LOS. We look at this question from two
angles. First we present a phase structure function analysis of the
dispersion measure data and estimate a power-law index of the electron
density spectrum.  Then we estimate the probability that such a
spectrum would produce a 22-y realization that was so strongly
dominated by the large, monotonic changes mentioned above.

We write the power spectrum of electron-density fluctuations as
\begin{equation}
P(q)\;=\; C_n^2\;q^{-\beta},\;\;\;\;\;\;\;\;\;\;\;\;[\qo<q<\qi]
\end{equation}
\noindent
where $\beta$ is the power law index, $\qo$ and $\qi$ are the spatial
frequencies corresponding to the outer and the inner boundary scale,
within which this power law description is valid. $C_n^2$ is the
amplitude, or strength, of the fluctuations. A quantity that is
closely related to the density spectrum which can be quantified by
observable variables is the phase structure function, $D_{\phi}(b)$,
with $b=2\pi/q$. This is defined as the mean square geometric phase
between two straight line paths to the observer, with a separating
distance of $b$ between them in the plane normal to the observer's
sight line. The phase structure function and the density power
spectrum are related by a transform (Rickett 1990; Armstrong,
Rickett, Spangler 1995),
\begin{eqnarray}
D_{\phi}(b)\;=\;\int_0^{\infty} 8\pi^2\lambda^2 r_e^2\;dz'\; 
\int_0^{\infty}q\;[1-J_0(bqz'/z)]\;dq \nonumber \\
\times P(q=0)
\end{eqnarray}
Here, $r_e$ is the classical electron radius (2.82$\times$10$^{-15}$ 
meters), $J_{\circ}$ is the Bessel function. Under the conditions 
that we have assumed, $D_{\phi}(b)$ is also a
power law (Rickett 1990; Armstrong, Rickett \& Spangler 1995), given
by

\begin{equation}
D_{\phi}(b)\;=\;\left(\frac{b}{b_{\rm coh}}\right)^{\beta-2}
\end{equation}
\noindent
where $b_{\rm coh}$ is the coherence spatial scale. Dispersion measure
can be written as
\begin{equation}
DM\;=2.410\times 10^{-16}\, \left[\frac{(\nu_1^2-\nu_2^2)}{ \nu_1^2
\nu_2^2}\right]\; \left(\frac{\Delta\phi}{f}\right)\; {\rm pc\;
cm^{-3}},
\end{equation}
\noindent
where $\Delta\phi$ is the difference in the arrival phases
$(\phi_2-\phi_1)$ of the pulse at the two barycentric radio
frequencies (Hz) $\nu_1$ and $\nu_2$, with $f$ being the barycentric
rotation frequency (Hz) of the pulsar. With this linear relation between DM
and geometric phase difference, then structure function can be written
as (KTR94)
\begin{eqnarray}
D_{\phi}(b_{\circ})&=&\left( \frac{2\pi}{\nu} \frac{{\rm Hz}}{ 2.410
\times 10^{-16}\;{\rm pc\;cm^{-3}}}\right)^2 \nonumber\\
&& \times \langle [DM(b+b_{\circ}) - DM(b)]^2 \rangle.
\label{eqn:struct}
\end{eqnarray}
Here, the angular brackets indicate ensemble averaging. The
transformation between the spatial coordinate $b$ (and the spatial
delay $b_{\circ}$) and the time coordinate $t$ (or time delay $\tau$)
is simply given by $b=V_\perp t$, where $V_\perp$ is the transverse
velocity of the LOS across the effective scattering screen.


\begin{figure*}
\begin{center}
\epsfig{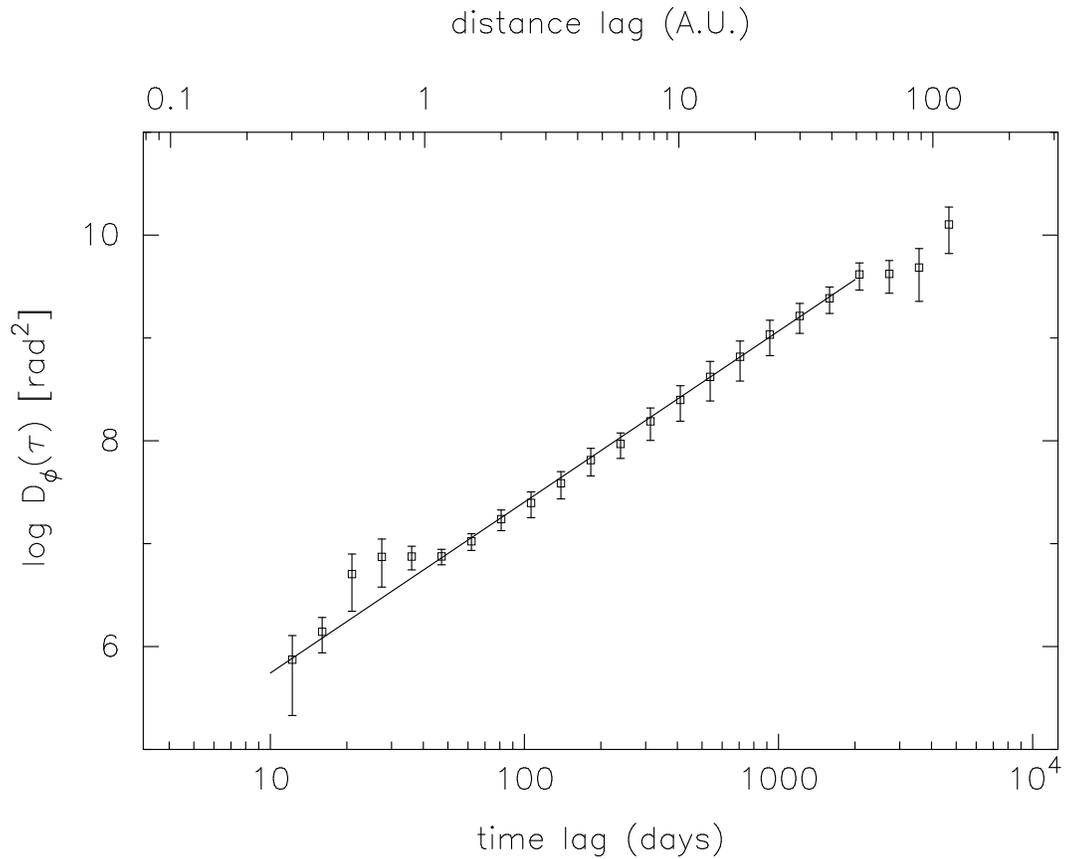}
\caption[]{The phase structure function derived with the help of
Equation \ref{eqn:struct} from the data displayed in Figure
\ref{fig:dmvstime}. Solid line represents the best fit for the data
in the time range of 30 days to 2000 days. For translating the time 
delay range into a space delay, we have assumed a sightline transverse 
velocity of 40 km sec$^{-1}$, which is half of the pulsar's peculiar 
velocity in its LSR. We have assumed an effective screen at a fractional 
distance of 0.5. See text for details.}
\label{fig:struct}
\end{center}
\end{figure*}

\begin{table}
\label{tab:scintparams}
\begin{center}
\caption[]{Measured and expected parameters.}
\begin{tabular}{c|c|c|c|c|c|c|c} \hline 
$\tausc$ & $\theta_H$ & $\tdiff$ & \multicolumn{2}{|c|}{$T_{\rm ref}$}
& \multicolumn{2}{|c|}{$m$} & {\bf $\nu$} \\ 
($\mu$s) & (mas) &
(s) & observed & expected & observed & expected & (MHz) \\
\hline\hline 
120$^{\dagger}$ & 14.6$^{b}$ & -- & 73 days & 3
y$^f$ & 0.33 & 0.14$^c$ & 327 \\ 
38$^e$ & -- & 100$^e$ & -- & -- & -- & -- & 430 \\
-- & -- & -- & 43.9 days & 6 mon$^g$ & 0.39 & 0.2 & 610 \\ 
-- & -- & 260$^a$ & 3 days$^d$ &
45 days$^g$ & 0.45 & 0.33$^c$ & 1400 \\ 
0.17$^e$ & -- & 444$^e$ & -- & -- & -- & -- & 1400 \\
\hline
\end{tabular}
\end{center}
$^{\dagger}$Has a time dependent RMS variation of 20$\mu$s\\
$^a$Cordes et al. (1986) \\
$^b$Gwinn et al. (1993) \\
$^c$Romani et al. (1986); Kaspi \& Stinebring (1992) \\
$^d$Lestrade et al. (1998) give the value as 13 days\\
$^e$Cordes et al. 1990 \\
$^f$Calculated with $\tref\sim\theta_HD/2\vlos$ \\
$^g$Extrapolated with $T_{\rm ref}\propto\lambda^{2.2}$
\end{table}

With the understanding that any difference in DM that we compute for a
time baseline from Figure \ref{fig:dmvstime} corresponds to a point in
the phase structure function, we can derive the phase structure
function on the basis of Equation \ref{eqn:struct}. This is given in
Figure \ref{fig:struct}.

There are several important points in Figure \ref{fig:struct}. The
solid line gives the best fit line for the data in the time interval
of 5 days to 2000 days. The derived values of the intercept and the
power law index ($\beta$) are,
\begin{eqnarray}
{\rm intercept} &=& 4.46\pm 0.09 \nonumber\\
\beta &=& 3.66\pm 0.04
\label{eq:fitpars}
\end{eqnarray}
The value of $\beta$ is remarkably close to the value expected from a
Kolmogorov power law distribution ($\beta=11/3$). We are using the
terminology ``intercept" only to indicate the value of
$\log[D_{\phi}(\tau)]$ when $\log[time lag (days)]$ is zero. Here, a
cautionary remark is warranted. Given the finite time span of our data
set, and the fact that the low spatial frequencies dominate the long
term variations in DM, we do not have a stationary sample of noise
spectrum. We have estimated the error in each bin of the structure
function as
\[
\sigma_s = \frac{\sigma_{D}}{\sqrt{N_i}},
\]
where $\sigma_{D}$ is the root mean square deviation with respect to
the mean phase structure function value in a bin, $D_{\phi}(\tau)$, 
and $N_i$ is the number of ``independent'' samples in the bin. This is 
estimated as the smaller of $(T/\tau)$ and the actual number of samples 
that have gone into the estimation of $D_{\phi}(\tau)$. 
Here, $T$ is the time span of the data.

By assuming that the transverse speed of the sightline across the
effective scattering screen is $\sim$40 km sec$^{-1}$ (half of
pulsar's velocity in its LSR), we can translate the delay range
between which this slope is valid, to 0.2 to 50 A.U.


The time delay value corresponding to the phase structure function
value of unity is, by definition, the coherent diffractive time scale
($\tdiff$) at the corresponding radio frequency, with the assumption
that the scattering material is uniformly distributed along the
LOS. From the fit parameters given in Equation~\ref{eq:fitpars}, 
this delay is 180 seconds. This should
be compared with the measured $\tdiff$ value of 444$ \pm $28 seconds
tabulated in Table 3. If we interpret the inner
scale cutoff value of $r_i \sim 1.3\times 10^9$ meters as the scale
size below which the slope ($\beta$) of the density fluctuation
spectrum changes to a value greater than that given in Equation
\ref{eq:fitpars}, then the fact that the measured $\tdiff$ value of
444 seconds being significantly greater than 180 seconds is
understandable. In the limiting case, where the slope of the 
density irregularity power spectrum changes to the
critical value of $\beta = 4$ below the inner scale cutoff value, the
expected $\tdiff$ value is about 1100 seconds. This makes it very
important to measure the exact frequency dependence of the diffractive
parameters like temporal scatter broadening and diffractive
scintillation time scale. To the best of our knowledge, Cordes et
al. (1990) show the most complete multi-frequency measurements of the
diffractive scintillation parameters of this pulsars. Their
measurements are not accurate enough to distinguish between such small
variations in slope.


While our analysis of DM variability suggests a Kolmogorov spectrum at
AU scales, we are struck by the long term monotonic decrease of DM and
wonder if we might be seeing the effects of smooth gradients in large
scale galactic structures that are not part of a turbulent cascade.
We performed a Monte Carlo simulation to investigate this.  In each
realization of the simulation, we generated with a different random
number seed, a screen of density fluctuations. We assumed that the
random fluctuations corresponding to a given spatial frequency are
described by a Gaussian function, but the total power as a function of
spatial frequencies is described by a single power law of index
--11/3.  Assuming that the screen is located at the mid point along
the sight line, we let the pulsar drift with its transverse velocity,
and measured the implied column density (DM) as a function of time.

We developed a procedure similar to that of Deshpande (2000) to
compare the observed $\dmt$ curve with the simulated ones.  From the
observed $\dmt$ curve, we computed the parameter $\Delta DM\;=\;[DM(t)
  - DM(t-\tau_{\circ})]$, where $\tau_{\circ}$ is the time delay. Our
aim is to compare the distribution of this parameter in very short
delays and very large delays. As we can see in Figure
\ref{fig:struct}, the structure function describes a well defined
slope between the delay range of $\sim$30 days to $\sim$2000 days. We
defined two delay bins, 30--60 and 1300--2000 days, within which we
monitored the distribution function of the quantity $\Delta DM$. From
this, we could infer that the distribution at the bin of 1300--2000
days had a span of $\sim 20 \sigma_s$, where $\sigma_s$ is the RMS
deviation of the distribution at the delay range of 30--60 days. That
is, the $\Delta DM$ values that we see at largest delays is as high as
20 times that of the typical deviations at short delays.  We performed
the same procedure on the simulated set of data to quantify the
likelihood of such deviations. We simulated 1024 number phase
screens. Out of these 1024 screens, we found that such large
deviations were possible $\sim$7\% of the times. This is perhaps not
surprising, as with such a steep spectrum, it is obvious that most of
the power is in large scales (smaller spatial frequencies), and hence
they tend to dominate our $\Delta DM$ measurements. We conclude that
while monotonic changes of this magnitude are rare, it is consistent 
with a turbulent cascade spectrum of density fluctutations.

\section{Frequency dependence of DM}
\label{sec-dmfreq}

Dispersion depends on the column density of electrons. In a
uniform medium radio wave propagation senses the average density
in a tube whose beam waist is set by the Fresnel radius
$\sqrt{z(1-z)\lambda D}$. In a
turbulent medium frequency-dependent multipath propagation can expand
this tube considerably.  With refraction, the center of the tube
wanders from the geometric LOS.  Indeed there may be a number of wave
propagation tubes, each with their independent relative gain. The
consequence is that DM and related effects will show frequency
dependence:
\begin{enumerate}
\item the effective DM depends on frequency.
\item the DM variations at lower frequencies will be
much ``smoother'' than that at higher frequencies, as the apparent
angular size of the source acts as a {\it smoothing function} on the
measured DM variations.
\item since the apparent size of the source is larger at low
frequencies, some features of the ISM that are visible at lower
frequencies may be invisible at higher frequencies!
\end{enumerate}

\begin{figure*}
\begin{center}
\epsfig{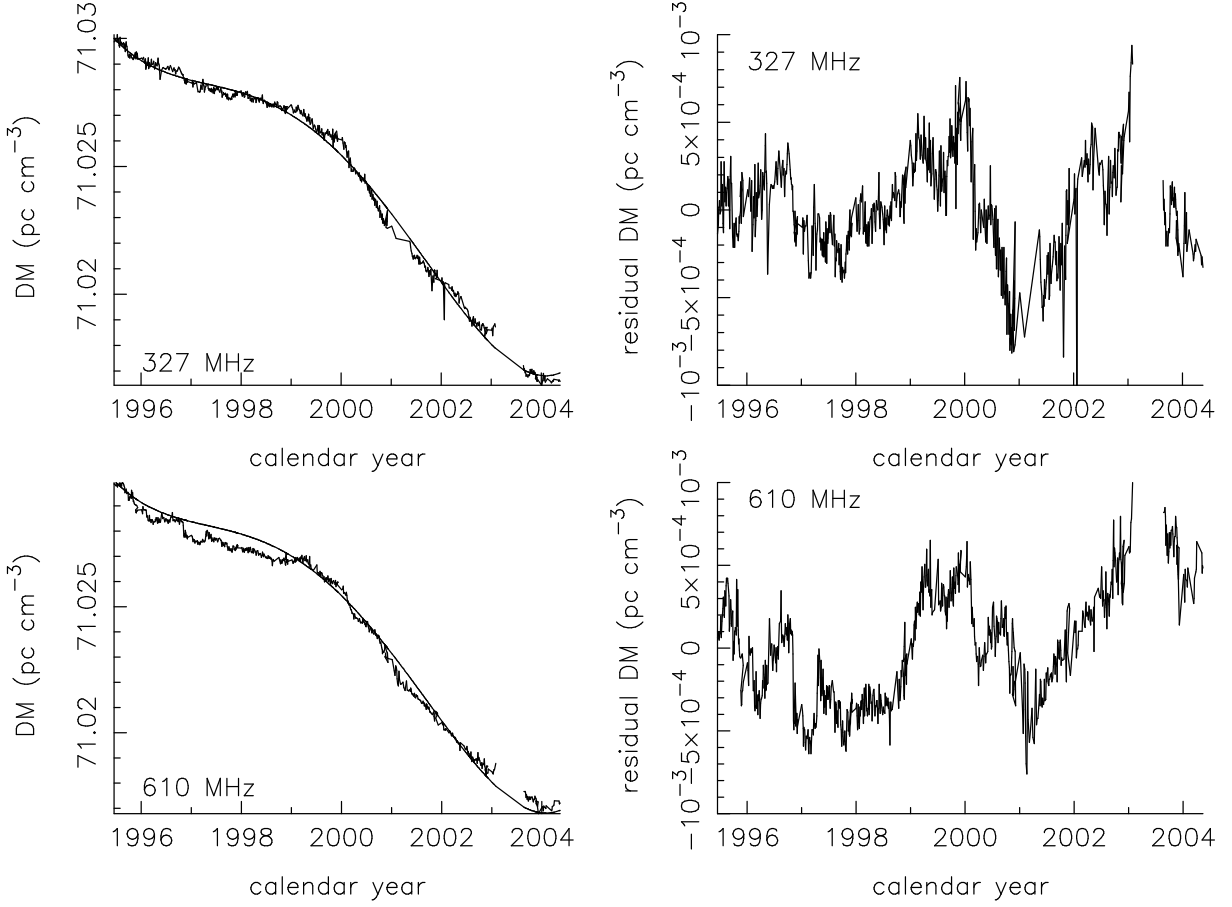}
\caption[]{Dispersion Measure variations at 327 and 610 MHz. The two
left hand side panels give DM as a function of time. The 327 MHz best
fit line, produced by a fourth order polynomial fit, is given as a
solid line in both 327 and 610 MHz plots. The residual DM, which is the
difference between the actual DM and the best fit line, is given in
the right hand side panels. See text for details.}
\label{fig:dmindiv}
\end{center}
\end{figure*}

We can explore these effects by assuming that the timing residuals at
327 MHz and 610 MHz, which are relative to the timing model derived at
higher frequencies that included removal of DM variations, are due to
DM variations.
The smoothing  effect of  scattering could be  revealed by  a spectral
analysis. The slow variations  were removed to pre-whiten the spectrum
that would otherwise  be severely contaminated.  The 327  MHz data was
fit to  a fourth order polynomial  and the result  was subtracted from
both data sets. The two right  side panels give the residual DM values
after subtracting  the best  fit curve from  the actual DM  curve. The
resulting  spectral  comparison fails  to  have  sufficient signal  to
clearly  demonstrate increased smoothing  at 327  MHz relative  to 610
MHz. Higher signal-to-noise ratio is required. The DM variations 
relative to long term trends  in  the  right-hand  panels of  Figure 
\ref{fig:dmindiv}  are different.

An important source of systematic error that can affect our analysis
here is the effect of scattering on the derived DM as a function of
time at a given frequency. At 327 MHz, as we described in
\S\ref{sec-anal}, we perform an elaborate procedure to fit for the
scatter broadening of the pulse profile, in order to compute the
``true'' TOA of the profile. However, we do not follow this procedure
at 610 MHz (or any other higher frequency). The error due to this can
be quantified easily from Figure \ref{fig:tausc}. The temporal scatter
broadening value varies by an RMS value of some 19.6 $\mu$s. With the
wavelength dependence of $\tausc\propto\lambda^{4.4}$, the expected
RMS variation at 610 MHz is 1.3 $\mu$s. The equivalent DM perturbation
at 610 MHz with respect to infinite frequency is $\sim
10^{-4}$~pc~cm$^{-3}$.



\section{Achievable timing accuracy}
\label{sec-timingerror}
In this section, we will estimate quantitatively errors introduced by
various scintillation related effects. For PSR B1937+21, although a
typical observation with highly sensitive telescopes like Arecibo
telescope helps us achieve a TOA accuracy of a few tens of
nanoseconds, the ultimate long term accuracy seems to be far greater
than this. In general, it is a combination of frequency independent
``intrinsic timing nose" from the pulsar itself, and the frequency
dependent effects, such as what we are addressing here. With some 18
years of data at 800, 1400 and 2200 MHz, Lommen (2002) quantifies the
timing timing residual, after fitting for position, proper motion,
rotation frequency and its time derivative (see also Kaspi et
al. 1994). A large fraction of the left over residuals is presumably
the intrinsic timing noise. As we have mentioned before, we have
absorbed a good part of this by fitting for the second time derivative
of the rotation frequency, $\ddot{f}$ (see
Table~2). In this section, our aim is to quantify
possible timing errors from various ``chromatic'' effects related to
interstellar scintillation.

\subsection{Fluctuation of apparent angular size}
The temporal variability of pulse broadening, $\tausc$, 
(as shown in Figure \ref{fig:tausc}) means that even the apparent
angular broadening of the source, $\theta_H$, is also changing as a
function of time. Since $\tausc \propto \theta_H^2$, with the RMS
variation in $\tausc$ of 19.6 $\mu$s at the radio frequency of 327
MHz, the corresponding variation in $\theta_H$ comes out to be
$\sim$8\% of the mean value. This change occurs with typical time
scales of $\sim$67 days, which is the time scale with which $\tausc$
changes.  Since we have only one epoch of $\theta_H$ measurement, we
have no way of observationally verifying the mean value or the time
scale of its variation.

\subsection{Image wandering and the associated timing error}
Due to non-diffractive scintillation that ``steers'' the direction of
rays (``refractive focussing''), the position of the pulsar is
expected to change as a function of time. This is an important and
significant effect, as it introduces a TOA residual as a function of
time, depending on the instantaneous position of source on the
sky. Several authors have investigated this effect in the past
(Cordes et al. 1986; Romani et al. 1986; Rickett \&
Coles 1988; Fey \& Mutel 93; Lazio \& Fey 2001). For a Kolmogorov
spectrum of irregularities ($\beta = 11/3$) with infinitismally small
inner scale cutoff, Cordes et al. (1986) predict the value of 
RMS image wandering as
\begin{equation}
\langle\delta\theta^2\rangle^{1/2} = 0.18\; {\rm mas} \; \left(
\frac{D_{\rm kpc}}{\lambda_{\rm cm}} \right)^{-1/6} \theta_H^{2/3}
\end{equation}
For an assumed distance to PSR B1937+21 of 3.6 kpc, this comes to 2
mas at 327 MHz (wavelength, $\lambda = 92$ cm). The value of 2 mas is
still significantly less than the apparent angular size of the source,
14.6 mas, measured by Gwinn et al. (1993). However, for a spectrum
with a steeper slope or with a significantly larger inner scale cutoff
(as in our case), the value of $\langle\delta\theta^2\rangle^{1/2}$ is
expected to be much larger, perhaps comparable to the value of
$\theta_H$.

In order to estimate the timing errors introduced by this image
wandering, we need an estimate of scattering measure ($SM$) and $C_n^2$
along the LOS to this pulsar. Following Cordes et al. (1991),
\begin{eqnarray}
SM &=& \int_0^D C_n^2(x)\;\;{\rm d}x \nonumber\\ &=&
\left(\frac{\theta_H}{128\;{\rm mas}} \right)^{5/3} \nu_{\rm
GHz}^{11/3} \nonumber\\ &=& 292\; \left( \frac{\tausc}{D_{\rm kpc}}
\right)^{5/6} \nu_{\rm GHz}^{11/3}.
\end{eqnarray}
Here, $\tausc$ is specified in seconds. $SM$ is specified in units of
kpc m$^{-20/3}$. Assuming a distance of 3.6 kpc, $\tausc$ = 120 $
\mu $s, $\nu$=0.327 GHz, the value of $SM$ comes to $\sim 8.8\times
10^{-4}$ kpc m$^{-20/3}$. Assuming that the scattering material is
uniformly distributed along the LOS, $C_n^2\sim 2.4\times 10^{-4}$
m$^{-20/3}$.

Then, for a Kolmogorov spectrum, the RMS timing residual due to the
image wandering can be written as (Cordes et al. 1986)
\begin{equation}
\sigma_{\delta t_{\theta}} = 26.5 \;\; {\rm ns} \;\; \nu^{-49/15}
D^{2/3} \left( \frac{C_n^2}{10^{-4} m^{-20/3}} \right)^{4/5}.
\end{equation}
With the computed value of $C_n^2$ and a distance of 3.6 kpc, this
amounts to 4.8 $\mu$s at 327 MHz. Given the frequency dependence, this
effect can be minimized by timing the pulsar at higher frequencies.
For instance, at frequencies of 1 GHz and 2.2 GHz, this error
translates to 125 and 2 nanosec, respectively. However, given the
significantly large value of the inner scale cutoff, the RMS timing
error that we have computed may well be a lower limit, and it is
likely to be higher.

Given the fact that the exact source position due to this effect is
unknown at any given time, it is very difficult to compensate for this
effect.

\subsection{Positional errors in solar system barycentric corrections}

As we saw above, due to image wandering, the apparent position of the
source wanders in the sky. This introduces yet another timing
error. While translating the TOA at the observatory to the solar
system barycenter, we assume a position which is away from the actual
apparent position at the time of observation. This introduces an
error, which can be quantified as (Foster \& Cordes 1990)
\begin{equation}
\Delta t_{\rm bary}\;=\; \frac{1}{c}\; (\vec{r_e}\cdot\hat{n}) \;
(1-z) \Delta\theta_r(\lambda).
\end{equation}
Here, $c$ is the velocity of light, the dot product term gives the
projected extra path length travelled by the ray due to Earth's annual
cycle around the Sun, and $\Delta\theta_r(\lambda)$ is the positional
error due to image wandering. Of course, this term is a function of
frequency, and hence the error accumulated is different at different
frequencies.

At 327 MHz, with an RMS image wandering angle of 2 mas, for an object
at the ecliptic plane, $\Delta t_{\rm bary}\sim 2$ $\mu$s. For PSR
B1937+21, this error amounts to $\sim$0.8 $\mu$s. At frequencies 1 
GHz and 2.2 GHz, this error translates to 85 and 17 nanosec, respectively.

\subsection{Timing error due to DM variation as a function of frequency}
\label{sec-accuracy}

An important issue that arises due to the frequency dependent DM
variation is the timing accuracy. Pulsars like PSR B1937+21 are known
for the accuracy to which one can compute the pulse TOA. Given this,
one wishes to eliminate any error that is incurred due to systematic
effects like what we have here. Between 327 and 610 MHz (the two
curves in Figure \ref{fig:dmindiv}), the typical relative fluctuation
of DM that we see is about $5\times 10^{-4}$ pc cm$^{-3}$. As we
discussed before, this is purely due to the fact that the effective
interstellar column length sampled at one frequency is different from
that at another frequency, due to the scatter broadening of the
source. At 610 MHz, this relative DM fluctuation corresponds to some 6
$\mu$s at 610 MHz. That is, at 610 MHz, typically an unaccounted
residual of 6 $\mu$s is incurred due to just effective DM errors. Even
if the behavior of the pulse emission is extremely stable, at low
frequencies, interstellar scattering limits our timing capabilities.

Due to the fact that dispersion delay goes as $\nu^{-2}$, although the
above mentioned effect seems significant, one should be able to reduce
it by going to higher frequencies. For instance, at 2.2 GHz, the
DM-limited TOA error for PSR B1937+21 will be $\sim$0.5$\mu$s. This
is not necessarily encouraging news, as a timing residual error of
0.5$\mu$s is large when compared to the accuracy that we can achieve
in quantifying the TOAs (a few tens of ns) for this pulsar, given
our observations with sensitive telescopes like Arecibo.

To summarize, although one takes into account time dependent DM
changes while analysing the data, in order to achieve high accuracy
timing, it is important to correct for a frequency dependent DM
term. This introduces another dimension of correction in the timing
analysis.

\section{Concluding remarks}
We have presented in this paper a summary of over twenty years of
timing of PSR B1937+21. These observations have been done over
frequencies ranging from 327 MHz to 2.2 GHz with three different
telescopes.

Given the agreement between the measured apparent angular broadening
and that estimated by the temporal broadening, and the measured
proper motion velocity and that estimated by the knowledge of
scintillation parameters, we conclude that the scattering material
is uniformly distributed along the sightline.

There are three significant discrepancies between the expected values
and the measured refractive parameters. These are,
\begin{enumerate}
\item The measured flux variation time scale is about an order of
magnitude shorter than what is expected from the knowledge of the
observed apparent angular broadening.
\item The flux variation time scale is observed to be directly
proportional to the wavelength, whereas it is expected to vary as
proportional to $\lambda^{2.2}$ (for a Kolmogorov spectrum).
\item The flux modulation index is observed to have a wavelength
dependence that is much ``shallower" than the expected value.
\end{enumerate}
These three discrepancies consistently imply that the optics is
``caustic-dominated''. This would mean that the density irregularity
spectrum has a large inner scale cutoff, $1.3\times 10^9$ m.  Our
extrapolation of the phase structure function from the regime sampled
by DM variations to the diffractive regime seems to indicate that the
expected $\tdiff$ value is considerably shorter than the measured
value. This is in favor of the above conclusion. Accurate measurements
of frequency dependence of diffractive parameters is much needed.

In general, Millisecond pulsars are known for their timing stability.
Potentially, we may achieve adequate accuracy in timing some of these
pulsars to understand some of the most important questions related to
the gravitational background radiation, or the internal structure of
these neutron stars. However, our analysis here shows that
interstellar scattering could be an important and significant source
of timing error.  As we have shown, although PSR B1937+21 is known to
produce short term TOA errors as low as 10--20 nanosec with sensitive
observations, the long term error is far larger than this. After
fitting for $\ddot{f}$ (which absorbes most of the achromatic timing
noise), the best accuracy that we can achieve for this pulsar is
$0.9\;\mu$sec at 1.4 GHz, and about 0.5 $\mu$sec at 2.2 GHz (by one of
the authors, Andrea Lommen). It appears that almost all of this error
can be accounted for by various effects that we have discussed in
\S\ref{sec-timingerror}. In general, for millisecond pulsars with
substantial DM, even if achromatic timing noise is small, interstellar
medium may be a major source of timing noise.

\acknowledgements We thank M. Kramer for sharing the data from
the Effelsberg-Berkeley Pulsar Processor (EBPP), and S. Chatterjee and
W. Brisken for sharing their VLBA based proper motion and parallax
results of PSR B1937+21 prior to the publication. This work was in part
supported by the NSF grant, AST--9820662.

\end{document}